\documentclass[twoside]{article}
\usepackage{fleqn}
\usepackage{espcrc2}
\usepackage{amssymb}
\usepackage{amsmath}
\usepackage{graphicx}
\usepackage{latexsym}

\def\mq{m_{\rm q}}

\def\mqt{\widetilde{m}_{\rm q}}
\def\mlt{\widetilde{m}_{\rm l}}
\def\mst{\widetilde{m}_{\rm s}}

\def\mqvalt{\widetilde{m}_{\rm q}^{\rm val}}

\def\mps{m_{\rm PS}}
\def\mv{m_{\rm V}}
\def\mK{m_{\rm K}}
\def\msbar{\overline{\rm MS}}
\def\cA{c_{\rm A}}
\def\bA{b_{\rm A}}
\def\bP{b_{\rm P}}
\def\bO{b_{\cal{O}}}
\def\ZA{Z_{\rm A}}
\def\ZS{Z_{\rm S}}
\def\ZP{Z_{\rm P}}
\def\ZO{Z_{\cal{O}}}
\def\Zmt{\widetilde{Z}_{\rm m}}
\def\csw{c_{\rm SW}}

\def\ksea{\kappa^{\rm sea}}
\def\klsea{\kappa_{\rm l}^{\rm sea}}

\def\kval{\kappa^{\rm val}}
\def\kvalc{\kappa_{\rm c}^{\rm val}}

\def\as{\alpha_{\rm s}}
\def\MeV{{\,\rm MeV}}
\def\GeV{{\,\rm GeV}}
\def\fm{{\,\rm fm}}
\def\Nf{N_{\rm f}}

\title{Light quark masses for dynamical, non-perturbatively O(a)
       improved Wilson fermions%
}
\author{D.~Pleiter
        \address{Deutsches Elektronen-Synchrotron DESY \& NIC,
                 D-15735 Zeuthen, Germany}
        \hspace{-0.25cm} $^,$\hspace{-0.15cm}
        \address{Institut f\"ur Theoretische Physik,
                 Freie Universit\"at Berlin, D-14195 Berlin, Germany},
        {\it QCDSF} and {\it UKQCD Collaboration}
}

\begin{document}

\begin{abstract}
We present results for the light quark masses in lattice QCD with two
degenerate flavours of dynamical fermions. We used configurations
generated by the {\it UKQCD} and {\it QCDSF} collaborations at
six different combinations of $\beta$ and $\ksea$.
\end{abstract}

\maketitle

\setcounter{footnote}{0}

\section{INTRODUCTION}

The calculation of the light quark masses is one of the major aims
of lattice gauge theories. In recent years these fundamental parameters
have been determined with a rather high accuracy using quenched QCD. At
least for the strange quark mass, systematic errors are believed to be
under control, except for quenching effects. However, for current
simulations of QCD with dynamical fermions this is not yet the case.
The main sources of uncertainty are the lack
of non-perturbative results for the renormalization constants and the
missing possibility to do a reliable continuum extrapolation.

For the calculations presented here we used dynamical configurations
with $\Nf=2$ flavours of degenerate quarks for six different combinations
of $\beta$ and $\ksea$. The coefficient for the improvement term of
the Wilson fermion action, $\csw(\beta)$, was taken from \cite{alpha98}.
The lattice spacing $a$ varies between 0.104 and 0.089
fm. For the lightest sea quarks the ratio $\mps/\mv$ was 0.6.
(For further details, see \cite{irving00}.)

We define the renormalized quark mass in the $\msbar$-scheme at scale
$\mu$ in the following way:
\begin{equation}
\mqt^{\msbar}(\mu) = \frac{(1 + \bA a \mq)\,\ZA^{\msbar}}
                          {(1 + \bP a \mq)\,\ZP^{\msbar}(\mu)}\;\mqt,
\end{equation}
\vspace*{-0.3cm}
where
\begin{align}
a \mq =\;& \frac{1}{2} \left(\frac{1}{\kappa} - \frac{1}{\kappa_c}\right) \\
\intertext{and}
a \mqt =\;& \frac{1}{2}  \frac{\langle \partial_4 A_4(t) P(0)\rangle}
                           {\langle P(t)   P(0)\rangle}
\end{align}
are the bare quark masses from the conserved vector current ({\it CVC}$\;$) and
the partially conserved conserved axial vector current ({\it PCAC}$\;$),
respectively.
In order to eliminate $O(a)$ effects introduced by the axial vector
current $A_\mu$, one has to replace $A_4$ in the last equation by
\begin{equation}
A_4 \rightarrow A_4 + c_A a \partial_4 P,
\end{equation}
where $P$ is the pseudoscalar density.
We therefore have to compute the following ratios on the lattice:
\begin{eqnarray}
a \mqt &=& a \mqt^{(1)} + \cA\; a \mqt^{(2)} \\
&=& \frac{\langle \partial_4 A_4(t) P(0)\rangle}
         {2 \langle P(t)   P(0)\rangle}
           + c_A \; \frac{\langle a \partial_4^2 A_4(t) P(0)\rangle}
                         {2 \langle P(t)   P(0)\rangle}.
\nonumber
\end{eqnarray}

Since both ratios can be measured rather precisely, we
are, for the rest of this talk, left with basically two problems:
\begin{itemize}
\item Determination of the renormalization constants $\ZO$, the
      coefficients $\bO$, which parametrize the quark mass dependence
      of the renormalization, and the improvement coefficient $\cA$.
\item Extrapolation of the lattice results to the physical quark masses.
\end{itemize}

In this talk we use $r_0 = 0.5\fm$ and take the lattice
results for $r_0/a$ from \cite{irving00} to set the scale.

\section{IMPROVEMENT AND RENORMALIZATION}

Until now, none of the renormalization constants and improvement coefficient,
except for $\ZA$ \cite{roger00}, have been calculated non-perturbatively
for $\Nf = 2$ improved fermions. We therefore will use perturbative results
and compare these with available non-perturbative numbers, e.g. quenched
results, in order to have some estimate of the possible errors.

Ordinary perturbation theory on the lattice is known to suffer from
problems because of large tadpole diagrams and the expansion in a
non-physical (bare) coupling constant. Usually it is better to apply
so-called tadpole improved perturbation theory, which refers to a
strategy to sum tadpole diagrams and to use a physical coupling
constant in the perturbative expansion.

The choice of the coupling constant to be used is not unique. We will
use the boosted coupling constant
$g^{*2} = g^2/u_0^4$
for bare quantities and the coupling constant in the $\msbar$-scheme
$(g^{\msbar})^2 = 4 \pi \; \as^{\msbar}(1/a)$
for the computation of renormalized quantities. $\as^{\msbar}(1/a)$ can
be calculated from the 4-loop expansion of the $\beta$-function
\cite{ritbergen97}, using $\Lambda^{\msbar}$ = 250 MeV.

\begin{figure}[h]
\vspace*{-0.4cm}
\begin{center}
\includegraphics[width=7.5cm]{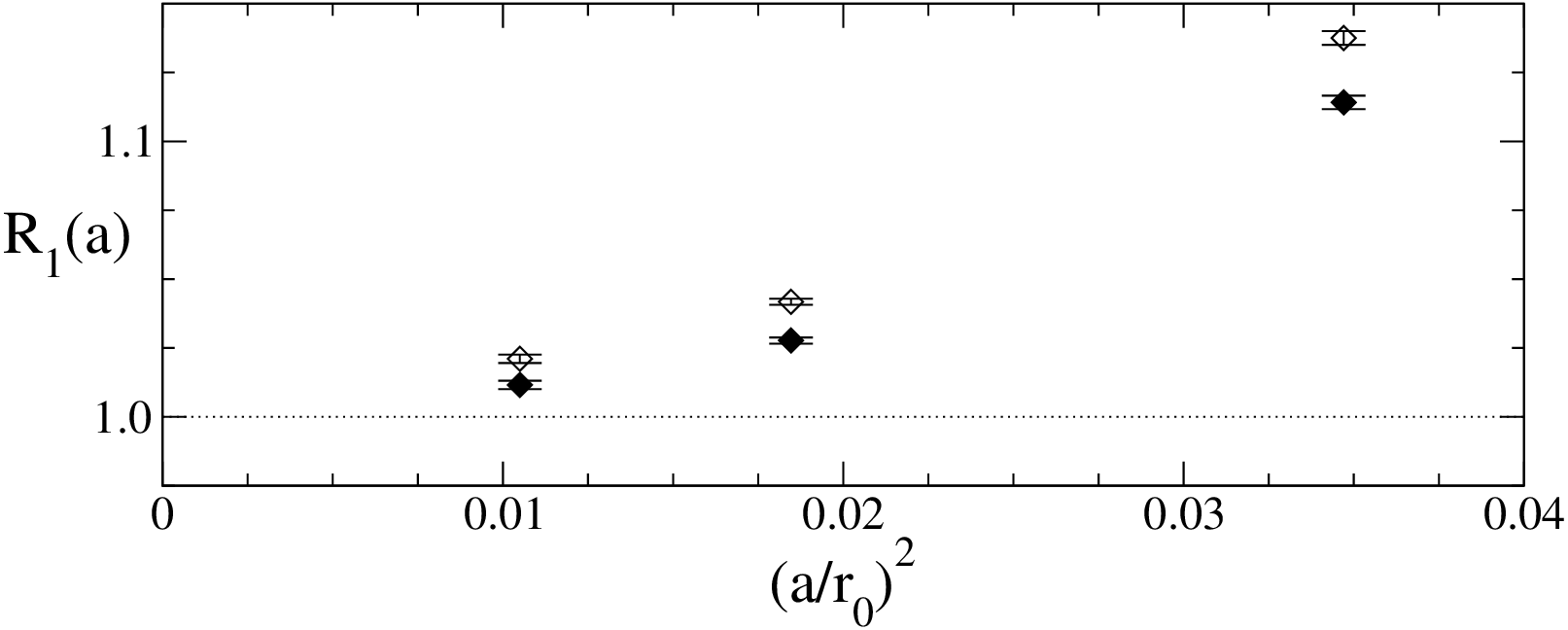}
\end{center}
\vspace*{-1.2cm}
\caption{\label{fig:r_ca} Ratio $R_1(a)$ defined in Eq.~(\ref{eq:r_ca})
         as a function of the lattice spacing using $\cA = 0$
         ($\Diamond$) and the tadpole-improved result ($\blacklozenge$).}
\vspace*{-0.3cm}
\end{figure}

The improvement coefficient $\cA$ is known from 1-loop perturbation
theory \cite{sint97}. In the quenched approximation $\cA$ has also
been computed non-perturbatively \cite{alpha97}. To parametrize the
difference between the non-perturbative and the tadpole-improved
results at the strange quark mass, it is convenient to define the ratio
\begin{equation}
R_1(a) = \left. \frac{\mqt^{(1)} + \cA \;\;\mqt^{(2)}}
                   {\mqt^{(1)} + \cA^{({\rm NP})} \mqt^{(2)}}
    \right|_{\mps = \sqrt{2} \mK}.
\label{eq:r_ca}
\end{equation}
In Fig.~\ref{fig:r_ca} we show $R_1(a)$ using our quenched results
\cite{roger99}. In the continuum limit we expect $R_1(0) = 1$.
Although $R_1(a)$ approaches this limit very quickly,
the results for the bare quark mass show discretization effects of
up to 15\% for $a \approx 0.1\fm$.

\begin{figure}[ht]
\begin{center}
\vspace*{-0.4cm}
\includegraphics[width=7.2cm]{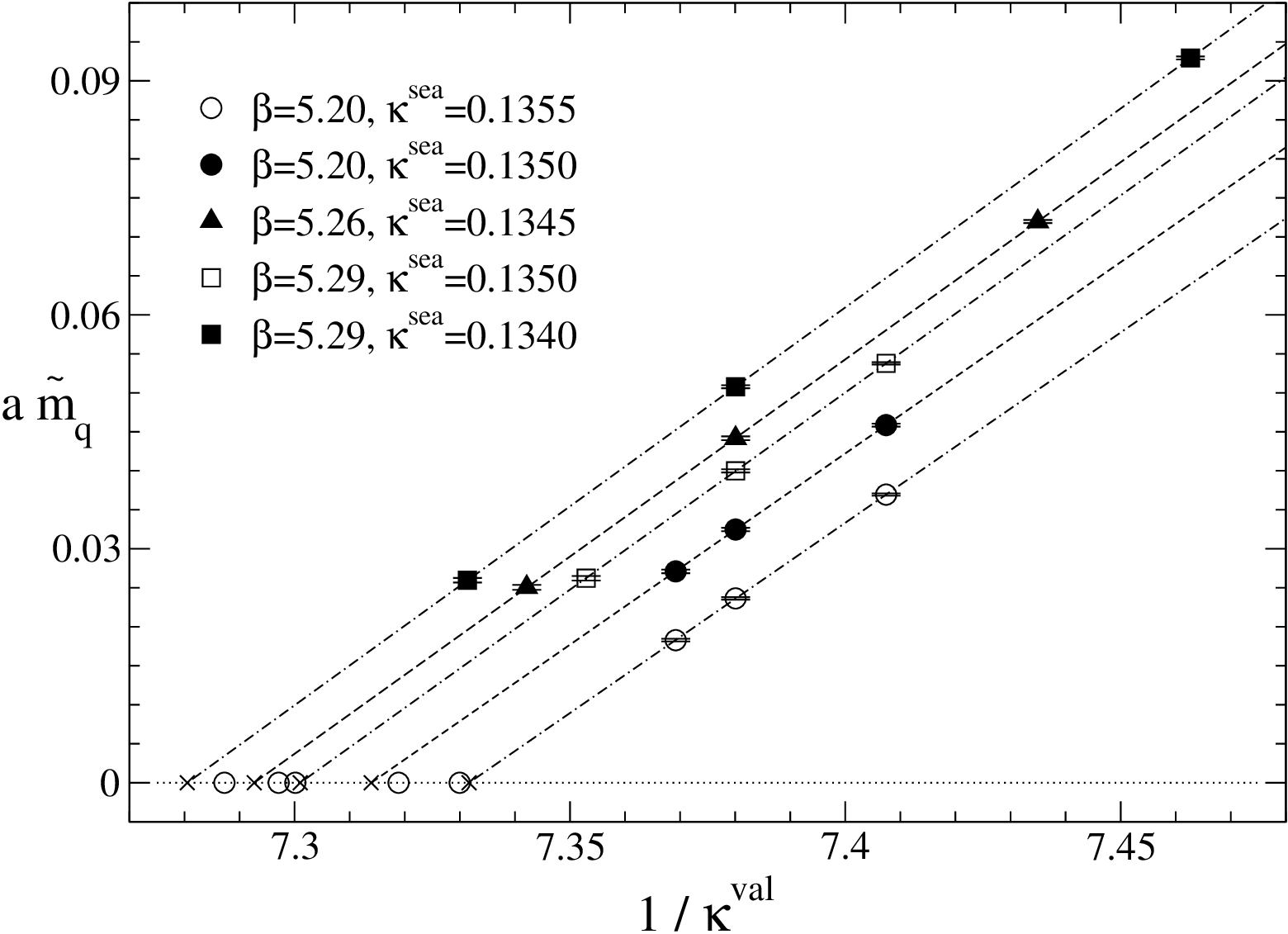}
\vspace*{-1.2cm}
\end{center}
\caption{\label{fig:kappac} Chiral extrapolation of $\mqt$ as function
         of $1/\kval$ for constant $(\beta,\ksea)$. The non-perturbative
         and the tadpole-improved results for $\kvalc$ are marked
         with $\times$ and $\circ$, respectively.}
\vspace*{-0.5cm}
\end{figure}

As a next example we compare the tadpole-improved result for
$\kvalc$, which is given to 1-loop tadpole-improved perturbation theory
by \cite{qcdsf96}
\begin{equation}
\begin{split}
\kvalc = \frac{1}{8}
 \big[1 + g^{*2} (& 0.025 - 0.029\;\csw u_0^3 \\
                  & - 0.012\;(\csw u_0^3)^2) \big] u_0^{-1},
\label{eq:kappac}
\end{split}
\end{equation}
with the non-perturbative results.
For each $(\beta,\ksea)$ we calculated $\mqt$ for three different
values of $\kval$. From a linear extrapolation
in $1/\kval$ to $\mqvalt = 0$ we determine $\kvalc$.
In Fig.~\ref{fig:kappac} we compare our non-perturbative results with
the results from Eq.~(\ref{eq:kappac}). In this case we find the relative
difference between both results to be less than 0.1\%.

\begin{figure}[t]
\begin{center}
\vspace*{0.2cm}
\includegraphics[width=7.2cm]{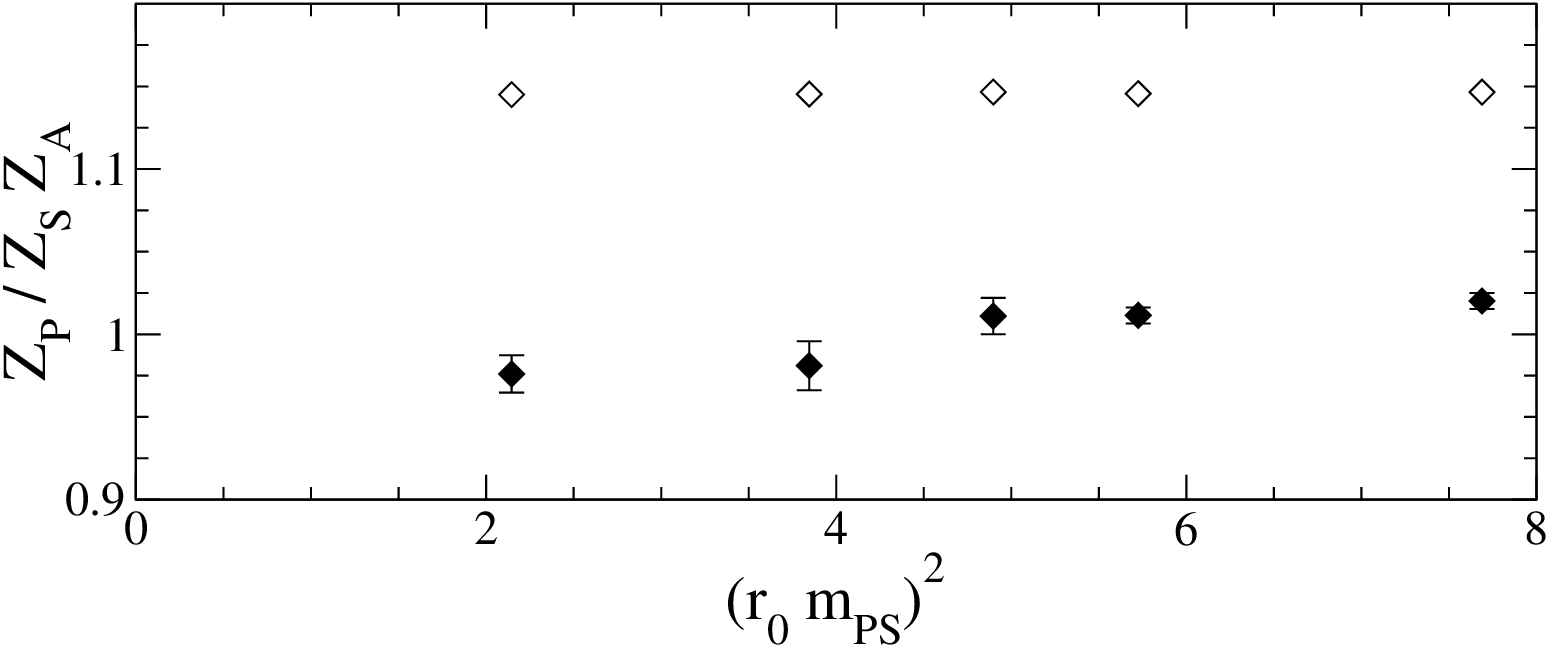}
\vspace*{-1.2cm}
\end{center}
\caption{\label{fig:ytwid} Non-perturbative ($\blacklozenge$) and
         tadpole-improved results ($\Diamond$) for $\ZP / \ZS \ZA$
         as a function of the pseudoscalar mass at $\ksea = \kval$.}
\vspace*{-0.5cm}
\end{figure}

%
The renormalization constants are known to 1-loop
tadpole-improved perturbation theory \cite{stefano97}.
$\ZP^{\msbar}(\mu)$ is scale dependent and we will use the
renormalization group equations, to scale our results to a convenient
scale, i.e.
\begin{equation}
\mqt^{\msbar}(2\GeV) =
  \frac{\Delta Z^{\msbar}(1/a)}{\Delta Z^{\msbar}(2\GeV)}\;
  \mqt^{\msbar}(1/a),
\end{equation}
where $\Delta Z(\mu)$ \cite{roger99}
is the factor which translates a scale and scheme
dependent value of the quark mass to its renormalization group invariant
value.

\begin{figure}[ht]
\begin{center}
\vspace*{-0.4cm}
\includegraphics[width=7.2cm]{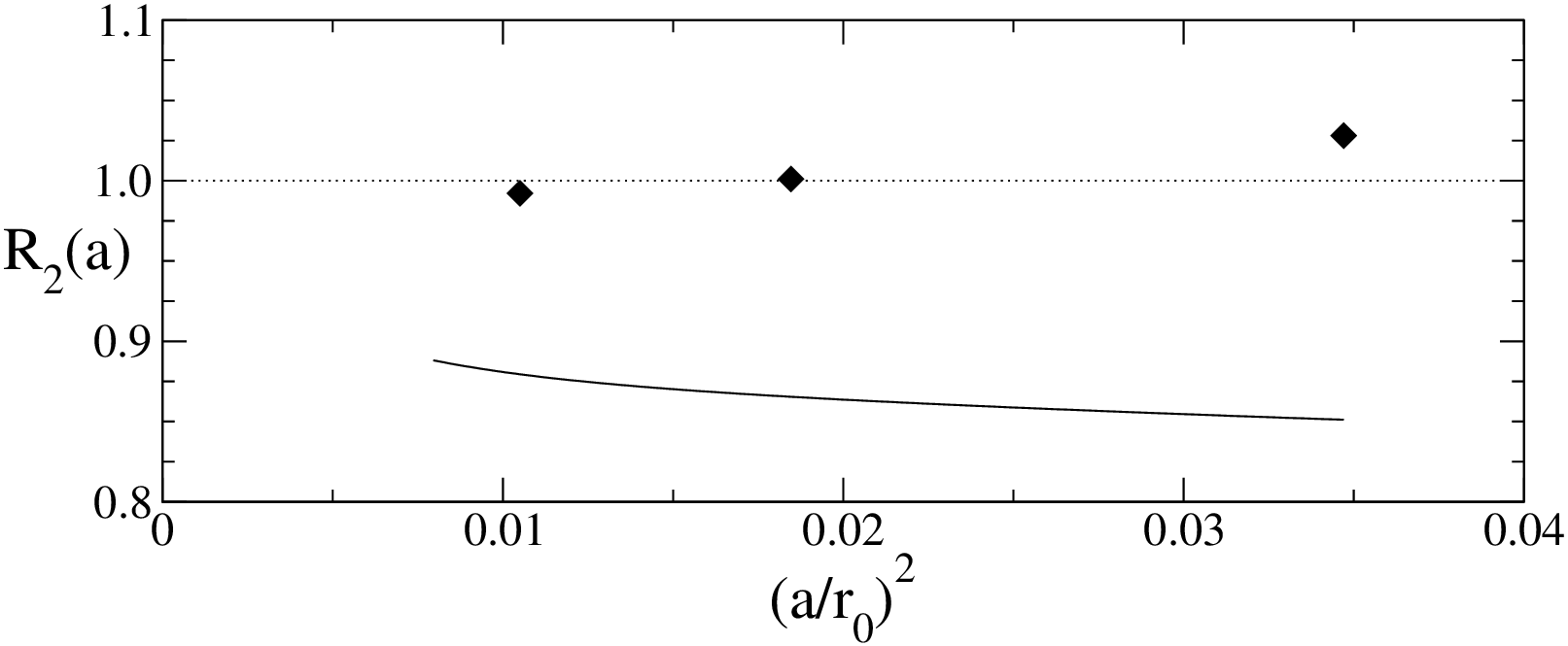}
\vspace*{-1.2cm}
\end{center}
\caption{\label{fig:r_Z} Ratio $R_2(a)$ defined in Eq.~(\ref{eq:r_Z})
         for quenched QCD at $\beta = 6.0, 6.2,$ and $6.4$ using
         results from ordinary (solid line) and tadpole-improved
         perturbation theory  ($\blacklozenge$).}
\vspace*{-0.6cm}
\end{figure}

To first order we expect that the following relation between $\mqt$
and $\mq$ holds:
\begin{equation}
\mqt = \frac{\ZP}{\ZS \ZA} \mq + O(\mq^2).
\end{equation}
From the slope of the fits shown in Fig.~\ref{fig:kappac} we get
a non-perturbative value for the ratio $\ZP / \ZS \ZA$.
In Fig.~\ref{fig:ytwid} we compare these results with the corresponding
tadpole-improved numbers. In this case we
find the difference to be of the order of 10\%. However, one might
hope that the results for $\Zmt(\mu) = \ZA / \ZP(\mu)$ from
tadpole-improved perturbation theory to be more reliable, since the
contributions from the tadpole diagrams cancel. For the quenched approximation
this can be checked, since $\Zmt(\mu)$ has
been calculated non-perturbatively in the Schr\"{o}dinger functional
(SF) scheme~\cite{alphaZ}. In order to compare perturbative and
non-perturbative results we looked at the ratio
\vspace*{-0.2cm}
\begin{equation}
R_2(a) = \frac{\Delta Z^{\msbar}(1/a) \Zmt^{\,\msbar}(1/a)}
              {\Delta Z^{\rm SF}(1/L) \Zmt^{\rm SF}(1/L)},
\label{eq:r_Z}
\end{equation}
where we expect $R_2(a) = 1 + O(a)$. The results for $R_2(a)$ are shown
in Fig.~\ref{fig:r_Z}.

In perturbation theory we have $\bA \approx \bP$. This has recently
been confirmed non-perturbatively for quenched QCD~\cite{alpha00}.
It is therefore expected that these coefficients will have only a small
effect on the results.

\section{QUARK MASSES}

\begin{figure}[t]
\begin{center}
\vspace*{0.2cm}
\includegraphics[width=7.2cm]{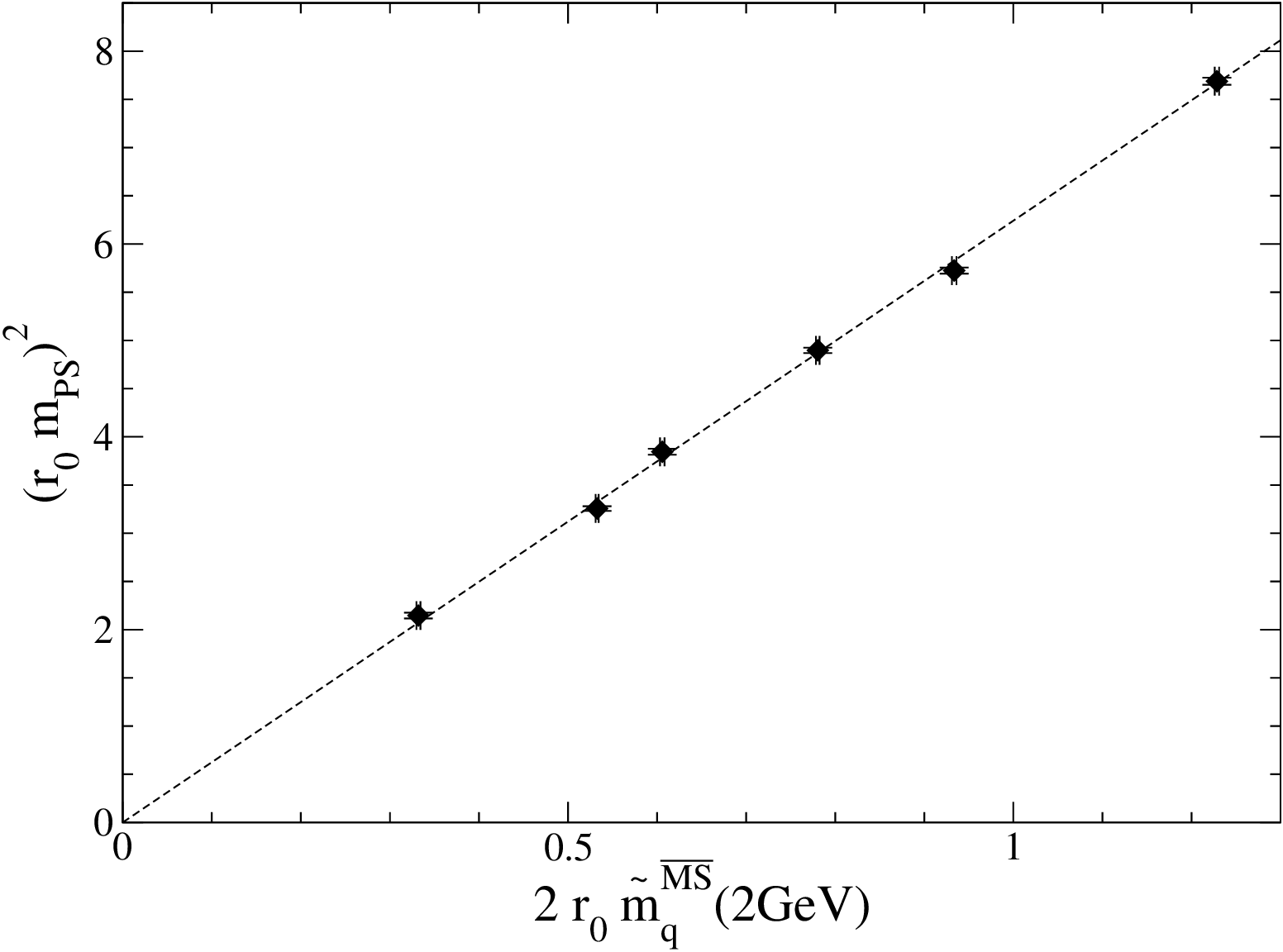}
\vspace*{-1.5cm}
\end{center}
\caption{\label{fig:mq.vs.mps} Renormalized quark mass as function of the
         squared pseudoscalar mass in units of $r_0$.}
\vspace*{-0.5cm}
\end{figure}

We are now able to combine our results and to calculate
$\mqt^{\msbar}(2\GeV)$.
In Fig.~\ref{fig:mq.vs.mps} we show the squared pseudoscalar mass
as a function of the renormalized quark mass at $\kval=\ksea$.

Even for large quark masses the
deviation from linearity is very small. Making a linear
ansatz we can compute the light and strange quark mass
by solving the equations
\begin{eqnarray}
b_1\;2 r_0 \mlt &:=& (r_0 m_{\pi}^{\rm phys})^2, \\
b_1\;(r_0 \mlt + r_0 \mst) &:=& (r_0 \mK^{\rm phys})^2.
\end{eqnarray}
Using the experimental values $m_{\pi}^{\rm phys} = 137.3\MeV$ and
$\mK^{\rm phys} = 495.7\MeV$ yields the results
$\mlt^{\msbar}(2\GeV) = 3.8(3)\MeV$ and $\mst^{\msbar}(2\GeV) = 96(4)\MeV$.

\begin{figure}[b]
\begin{center}
\vspace*{-0.7cm}
\includegraphics[width=7.2cm]{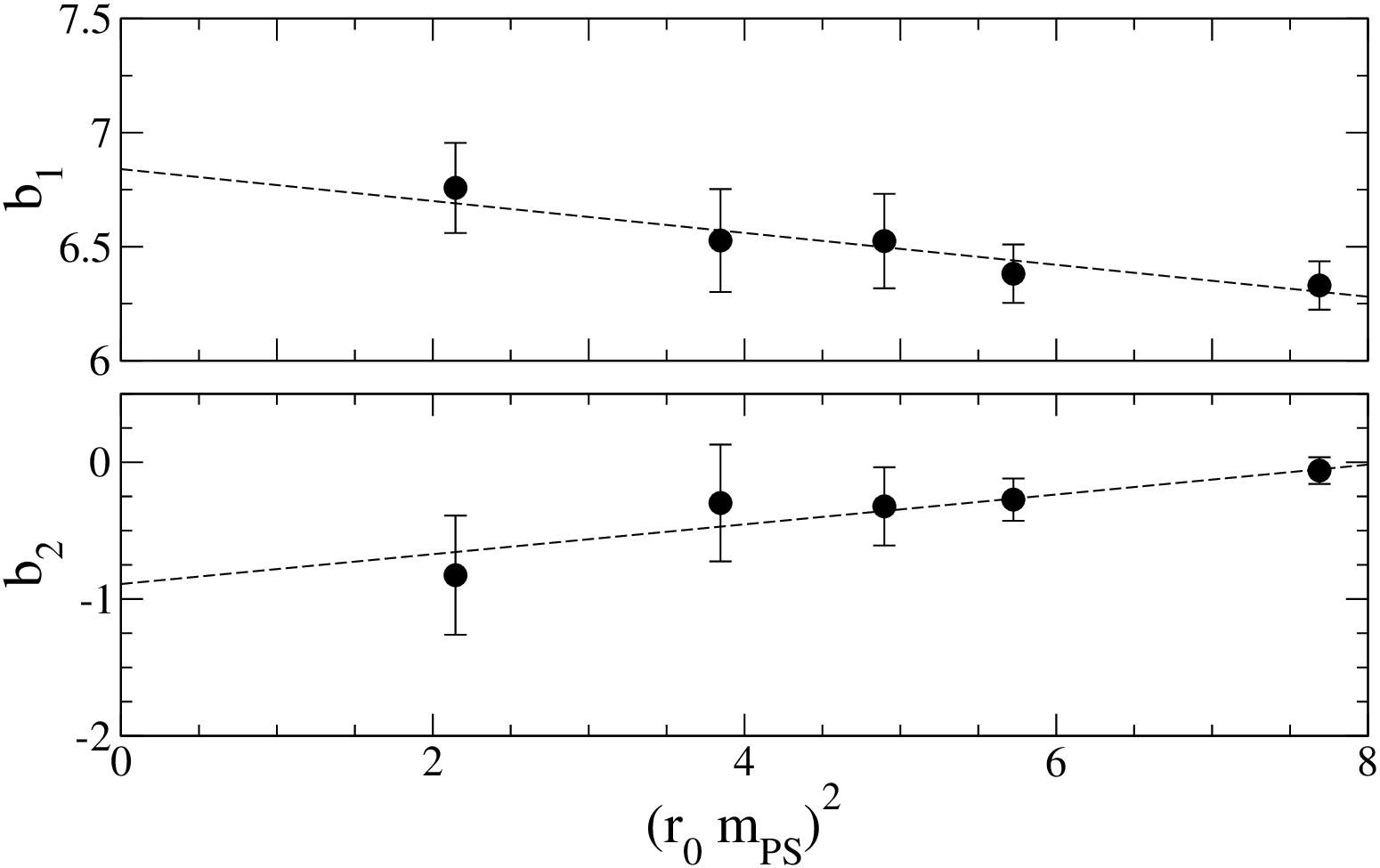}
\vspace*{-1.4cm}
\end{center}
\caption{\label{fig:b2} Coefficients $b_1$ and $b_2$ as a function of the
         squared pseudoscalar mass at $\ksea=\kval$.
         The lines show a linear fit.}
\end{figure}

For a more sophisticated ansatz we make use of partially quenched chiral
perturbation theory.
To 1-loop order the pseudoscalar mass can be written in the following
form \cite{sharpe99}:
\begin{equation}
\mps^2 = b_1(\ksea)\;2\mqvalt + b_2(\ksea)\;\left(2\mqvalt\right)^2
\end{equation}
%
The coefficients $b_1$ and $b_2$ can
be calculated from a quadratic fit to our results at constant
$(\beta,\ksea)$. The results, which show a rather mild dependency on
$\ksea$, are plotted in Fig.~\ref{fig:b2}.
%
To calculate the light and strange quark masses we use the ansatz
\begin{align}
(r_0 m_{\pi}^{\rm phys})^2 :=\;& b_1(\klsea)\;r_0\,2\mlt + 
                                 b_2(\klsea)\;(r_0\,2 \mlt)^2 \nonumber\\
(r_0 \mK^{\rm phys})^2     :=\;& b_1(\klsea)\;r_0 (\mlt+\mst) + \\\nonumber
                               & b_2(\klsea)\;[r_0 (\mlt+\mst)]^2.
\end{align}
We finally get the following results:
\begin{eqnarray}
\mlt^{\msbar}(2\GeV) &=& 3.5(2) \MeV, \\
\mst^{\msbar}(2\GeV) &=& 90(5) \MeV. \nonumber
\end{eqnarray}

In a previous quenched calculation~\cite{roger99}
we found $\mst^{\msbar}(2\GeV) = 105(4)\MeV$. Since systematic errors,
which are currently not under control, might be of the order 10\%, we
can not conclude that the (small) difference between these results
is a quenching effect.

\section{CONCLUSIONS}

We did a preliminary analysis of the light quark masses using
two flavours of degenerate, non-perturbatively improved fermions.
We used tadpole-improved perturbation theory, but tried to give
some idea of the systematic errors.
For the extrapolation to the physical region we used
partially quenched chiral perturbation theory.  The difference
between our final results and earlier published quenched results
was found to be small.

\vspace*{-0.1cm}

\section*{ACKNOWLEDGEMENTS}

The numerical calculations were performed on the Hitachi {SR8000} at
LRZ (Munich), the Cray {T3E} at EPCC (Edinburgh), NIC (J\"ulich) and
ZIB (Berlin) as well as the APE/Quadrics at DESY (Zeuthen).
We wish to thank all institutions for their support.
UKQCD acknowledges PPARC grants GR/L22744 and PPA/G/S/1998/000777.


\vspace*{-0.1cm}

\begin{thebibliography}{9}
\bibitem{alpha98}	K.~Jansen and R.~Sommer,
			Nucl. Phys. B530 (1998) 185
			(hep-lat/9803017).
\bibitem{irving00}	A.~Irving, this conference.
\bibitem{roger00}	R.~Horsley, this conference.
\bibitem{ritbergen97}	T.~van~Ritbergen, J.A.M.~Vermaseren, S.A. Larin,
			Phys. Lett. B400 (1997) 379.
\bibitem{sint97}	S.~Sint, P.~Weisz,
			Nucl. Phys. B502 (1997) 251
			(hep-lat/9704001).
\bibitem{alpha97}	M.~L\"{u}scher, S.~Sint, R.~Sommer, P.~Weisz, U.~Wolff,
			Nucl. Phys. B491 (1997) 323
			(hep-lat/9609035).
\bibitem{roger99}	QCDSF Collaboration,
			Phys. Rev. D62 (2000) 54504
			(hep-lat/9908005).
\bibitem{qcdsf96}	QCDSF Collaboration,
			Phys. Lett. B391 (1997) 388
			(hep-lat/9609008).
\bibitem{stefano97}	QCDSF Collaboration,
			Nucl. Phys (Proc. Suppl.) 63 (1998) 874
			(hep-lat/9709049);
			hep-lat/0007004, to appear Nucl. Phys.
\bibitem{alphaZ}	S.~Capitani, M.~L\"{u}scher, R.~Sommer, H.~Wittig,
			Nucl. Phys. B544 (1999) 669
			(hep-lat/9810063).
\bibitem{alpha00}	M.~Guagnelli, R.~Petronzio, J.~Rolf, S.~Sint,
			R.~Sommer, U.~Wolff
			(hep-lat/0009021).
\bibitem{sharpe99}	S.~Sharpe, N.~Shoresh,
			Nucl. Phys. (Proc. Suppl.) 83 (2000) 968
			(hep-lat/9909090).
\end{thebibliography}
\end{document}